\documentclass[aps,pra,preprint,showpacs,amsmath,amssymb]{revtex4-1}
\usepackage{color}
\usepackage{graphicx}
\usepackage{dcolumn}
\usepackage{bm}
\begin{document}

\title{Entanglement between electronic and vibrational degrees of freedom in a laser-driven molecular system}

\author{Mihaela Vatasescu}
\email[]{mihaela\_vatasescu@yahoo.com}
\affiliation{Institute of Space Sciences - INFLPR,
MG-23, 77125 Bucharest-Magurele, Romania}


\begin{abstract}

We investigate the entanglement between electronic and vibrational degrees of freedom
produced by a vibronic coupling  in a molecular system described in the Born-Oppenheimer
approximation. Entanglement in a pure state of the Hilbert space $\cal{H}$$=$$\cal{H}$$_{el}$$\bigotimes$$\cal{H}$$_{vib}$ is quantified using
the von Neumann entropy of the reduced density matrix and the reduced linear  entropy. 
Expressions for these entanglement measures are derived for the $2 \times N_v$ and $3 \times N_v$ cases of the bipartite entanglement, where 2 and 3 are the dimensions  of the electronic Hilbert space 
$\cal{H}$$_{el}$, and $N_v$ is the dimension of $\cal{H}$$_{vib}$.
We study the entanglement dynamics for two electronic states coupled by a laser pulse 
(a $2 \times N_v$ case), taking as an example a coupling between the $a^3\Sigma_{u}^{+} 
(6s,6s)$ and $1_g(6s,6p_{3/2})$ states  of the Cs$_2$ molecule.
The reduced linear entropy expression obtained for the
$3 \times N_v$ case is used to follow the entanglement evolution in a scheme proposed
for the control of the vibronic dynamics in a Cs$_2$ cold molecule, implying the
$a^3\Sigma_{u}^{+}(6s,6s)$, $0_g^-(6s,6p_{3/2})$, and $0_g^-(6s,5d)$ electronic states,
which are coupled by a non-adiabatic radial coupling and a sequence of chirped laser pulses.

\end{abstract}


\maketitle

\section{\label{sec:intro}Introduction}

Quantum entanglement, central to the foundations of quantum theory
\cite{4horodecki09}, is today a reference concept shaping the understanding  of various
quantum phenomena in physics, chemistry, and quantum biology. With the emergence of quantum
information theory entanglement was also recognized as a fundamental resource for quantum
computation and quantum communication \cite{bookNC}.

In the last twenty years atomic and molecular physics had a particularly
fortunate encounter with quantum information theory, sustained by the continuous development
of experimental techniques able to produce extremely controllable ultracold atomic and
molecular systems.  Entanglement has been explored  in a variety of experiments employing
highly controlled atomic systems like cold trapped ions \cite{wineland13}, Rydberg atoms
crossing a "photon box"  \cite{haroche13}, or neutral atoms in optical lattices
\cite{zoller99,*bloch08,*zoller12}.
The controlled creation of entanglement between pairs of atoms
trapped  in an optical lattice was used for precision measurements of atomic scattering
properties \cite{bloch04}, atomic spectroscopy using quantum logic was implemented with
trapped atomic  ions \cite{wineland05}, and quantum metrology was performed using "designer atoms"
\cite{blatt06}. A similar trend becomes increasingly possible in molecular physics, due to
the  progress in  the formation of ultracold molecules.  Proposals for molecular entanglement
creation are considering ultracold polar molecules \cite{demille02,*yelin09,*kbwhaley13} as
interesting systems for quantum information manipulation and promising platforms for quantum
computation.

In addition to these developments there is also an increased interest in using
quantum entanglement and quantum information concepts to describe the structure of atoms and
molecules and related phenomena. A recent review focusing on "essential entanglement for atomic
and molecular physics" \cite{tichy11} shows the specificity of this research program 
which considers physical objects far from the idealized systems familiar from the quantum
information science and for which the identification of subsystems which can carry entanglement
is non-trivial. Within this program various theoretical investigations have been
advanced, including studies of  entanglement in two-electron atomic systems \cite{osenda07,*coe10,
*manzano10,*dehesa10,*dehesa12,*benenti13,*lin13} and investigations of the entanglement between the
internal electronic and the external translational degrees of freedom of trapped atoms \cite{roghani12}.
Studies of entanglement in molecular systems have considered the
entanglement associated with the dissociation
of diatomic molecules \cite{kurizki01,*hornberger10,*esquivel11,*mshapiro12},
entanglement in Rydberg molecules \cite{matzkin06},
 and dynamical entanglement of vibrations in triatomic molecules \cite{hou06,*zhai13}.

Proposals for quantum computing using molecular internal degrees of freedom 
(electronic, vibrational, and rotational) have opened the door to stimulative research
in molecular systems driven by shaped light pulses, in which optimal quantum control theory is used to
find the driving fields which play the role of quantum logic gates \cite{lidar01,*tesch-riedle02,*palao-kosloff02,*vala02,*gollub06,*troppmann06,*brown06,*babikov07,*mishima08}. 
These developments have stimulated the interest in the characterization of entanglement in laser-driven molecular systems.

The present work investigates the entanglement between electronic and nuclear degrees of freedom
 in a molecule. Few studies have treated this subject and these works consider Hilbert spaces with low dimensionality. Special attention is attached to double-welled chemical
systems, as embodying electronic-vibrational entanglement through the role played by the wavefunction
delocalization \cite{sjoqvist00,mckemmish11}, and to entanglement in relation with quantum chaos induced
by non-adiabatic interaction due to the breakdown of the Born-Oppenheimer approximation \cite{fujisaki04}.

Here we consider a molecular system (diatomic molecule) described in the Born-Oppenheimer (BO) approximation
which separates the electronic and nuclear motion, leading to the factorization of the molecular
wavefunction into an electronic and a rotational-vibrational part. The rotational degree of freedom is
neglected, and, therefore, the system can be described by a Hilbert space which is a
tensor product $\cal{H}$$=$$\cal{H}$$_{el}$$\bigotimes$$\cal{H}$$_{vib}$ of electronic and vibrational Hilbert
spaces of finite dimensions. We will analyze the entanglement between electronic and vibrational degrees of
freedom produced by a coupling between electronic states: this coupling could be produced by an external source,
such as laser pulses acting on the molecular system, or it could be a non-adiabatic interaction neglected in the
Born-Oppenheimer approximation.
We consider pure states of the bipartite system (el $\bigotimes$ vib), and quantify the entanglement using
the von Neumann entropy of the reduced density matrix and the linear entropy related to the purity of the
reduced density matrix.  We derive formulas for these measures of entanglement, which can be employed to 
follow the entanglement evolution in relation to the intramolecular dynamics. 

We show results for the entanglement dynamics in two cases of temporal evolution in a laser-driven molecule.
A first example treats the case of a laser coupling between the $a^3\Sigma_{u}^{+}(6s,6s)$ and $1_g(6s,6p_{3/2})$ electronic states of the Cs$_2$ molecule (a $2 \times N_v$ case). 
The second example
follows the entanglement evolution quantified by the linear entropy (a $3 \times N_v$ case)
in a theoretical control scheme proposed 
to create Cs$_2$ vibrationally cold molecules 
using a multichannel tunneling observed in the cesium photoassociation spectrum.
The scheme 
employs the electronic states $a^3\Sigma_{u}^{+}(6s,6s)$, $0_g^-(6s,6p_{3/2})$, and $0_g^-(6s,5d)$ of the
Cs$_2$ molecule, which are coupled by a non-adiabatic radial coupling and a sequence of chirped laser pulses. 
In both cases the entanglement dynamics is analyzed
in relation to the characteristic times specific to the vibronic couplings
and intramolecular dynamics.

The structure of the paper is as follows. Sec.~\ref{sec:BOmolec} briefly reviews the theoretical framework
of the BO approximation. In Sec.~\ref{sec:2timesN}  the $2 \times N_v$ case of the bipartite entanglement
is studied and  expressions for the reduced von Neumann entropy and reduced linear entropy are derived.
The example of two electronic states coupled by a laser pulse is contained in Sec.~\ref{sec:dyn1g}.
Sec.~\ref{sec:3electr} treats the $3 \times N_v$ case, deducing the corresponding formula
for the reduced linear entropy.
Sec.~\ref{sec:3elchirp} follows the entanglement evolution quantified by the linear entropy in the theoretical
control scheme proposed to create Cs$_2$ vibrationally cold molecules using a sequence of chirped laser pulses. 
Sec.~\ref{sec:conclu} contains our final remarks.

\section{\label{sec:BOmolec} Molecular model: Born-Oppenheimer approximation and vibronic couplings between electronic states}

We briefly review some basic notions used in the description of a diatomic molecule
in the BO approximation \cite{bookLefebField}. The mass difference between nuclei and electrons justifies
the so-called clamped nuclei electronic Schr\"odinger equation, written for the electronic Hamiltonian
$H^{el}$ (i.e. the total molecular Hamiltonian without the nuclear kinetic-energy part):
\begin{equation}
H^{el}\phi_{n}^{el}(\vec{r_{i}};R)=U_{n}(R)\phi_{n}^{el}(\vec{r_{i}};R),
\end{equation}
where $R$ is the internuclear distance and $\{ \vec{r_{i}} \}$ the electronic coordinates expressed in
the molecule-fixed coordinate system. This equation produces the adiabatic potential-energy surfaces
$U_{n}(R)$ as eigenvalues of the electronic Hamiltonian and the electronic wavefunctions
$\phi_{n}^{el}(\vec{r_{i}};R)$, depending parametrically on R, as orthonormal eigenstates of $H^{el}$.

The molecular ro-vibronic wavefunction $\Psi_{mol}(\vec{R},\vec{r_{i}};t)$ can be expanded in
the basis set of the electronic eigenfunctions $\{ \phi_{n}^{el}(\vec{r_{i}};R) \}$:
\begin{equation}
  \Psi_{mol}(\vec{R},\vec{r_{i}};t) = \sum_{n}\psi_{n}(\vec{R};t)
  \phi_{n}^{el}(\vec{r_{i}};R).
  \label{eldevelop}
\end{equation}
In Eq.~(\ref{eldevelop}) we have introduced the time-dependent picture emphasizing that the temporal
dependence is contained in the nuclear wavefunctions $\psi_{n}(\vec{R};t)$. In a stationary picture,
the coefficients $\psi_{n}(\vec{R})$ depending on the nuclear geometry are the rotational-vibrational
wavefunctions. If the expansion (\ref{eldevelop}) is inserted in the Schr\"odinger equation of the
full molecular Hamiltonian, one obtains a system of coupled differential equations describing the motion
of the nuclei in an electronic state, coupled with the nuclear motion in all other electronic states.
In the BO approximation this coupling with other electronic states is neglected, providing a good
description if the electronic wavefunctions depend only weakly on $R$. Then, the sum in
Eq.~(\ref{eldevelop}) can be reduced to a single term:
\begin{equation}
  \Psi_{mol}(\vec{R},\vec{r_{i}};t) \approx \psi_{n}(\vec{R};t) \phi_{n}^{el}(\vec{r_{i}};R).
  \label{BOfactor}
\end{equation}
This factorization of the total wavefunction into an electronic and a rotational-vibrational part is
essential to the BO approximation. The consequence is that the nuclear motion in an electronic state
is uniquely determined by the corresponding electronic potential,  which allows one to write a
rotational-vibrational Schr\"odinger equation for each electronic state. As the radial and angular
variables of the nuclear motion can be also separated, the Schr\"odinger equation which defines the
rovibrational eigenfunctions $\chi^n_{v,J}(R)$ corresponding to an electronic potential $U_{n}(R)$
can take the form
\begin{eqnarray}
  \left(-\frac{\hbar^{2}}{2\mu}\frac{\partial^2}{\partial R^2}+
  U_{n}(R)+\frac{\hbar^{2}J(J+1)}{2 \mu R^{2}}\right) \chi^n_{v,J}(R)\nonumber\\
 = E^n_{v,J}\chi^n_{v,J}(R),
\label{eqSelvib}
\end{eqnarray}
where $\mu$ is the nuclear reduced mass, J quantifies the rotational angular momentum, and
$E^n_{v,J}$ are the energies of the rovibrational levels corresponding to the electronic state $n$.

Due to the fact that here the rotational degree of freedom is neglected  the molecular system is
described by the  Hilbert space  $\cal{H}$$=$$\cal{H}$$_{el}$$\bigotimes$$\cal{H}$$_{vib}$.
The molecular wavefunction corresponding to an electronic state $n$, BO factored into
electronic and vibrational wavefunctions, is
\begin{eqnarray}
  \Psi_{mol}^{n}(R,\vec{r_{i}},t)= \frac{1}{R} \chi_{n}(R,t) \phi_{n}^{el}(R;\vec{r_{i}})\nonumber\\
=\frac{1}{R} \left( \sum_{v} c^n_{v}(t) \chi^n_{v}(R) \right)  \phi_{n}^{el}(R;\vec{r_{i}}).
\label{elvibfactor}
\end{eqnarray}
In Eq.~(\ref{elvibfactor}) the vibrational wavepacket $\chi_{n}(R,t)$ is developed in the orthonormal
basis set of the vibrational eigenstates $ \{ \chi^n_{v}(R) \}$ corresponding to the electronic potential
$U_{n}(R)$. $\{ \chi^n_{v}(R) \}$ are solutions of Eq.~(\ref{eqSelvib}) for J=0 or for fixed J.

The BO states are generally those on which the molecular description is built
\footnote{\text{"}In order to go beyond the BO approximation, it is necessary to use a BO representation
\text{"}, as it is expressed in Ref.~\cite{bookLefebField}.}
by including various coupling mechanisms between the electronic states, such as non-BO coupling terms
subsequently introduced in the description, or vibronic couplings caused by external fields. In the
following we consider a molecular system which can be described by a pure state belonging to the bipartite
Hilbert space $\cal{H}$$_{el}$$\bigotimes$$\cal{H}$$_{vib}$ of finite dimension. Our aim is to derive
formulas for the entanglement between electronic and vibrational
degrees of freedom produced by vibronic couplings of the electronic states. In the next sections we will
explore the cases of bipartite entanglement in pure states belonging to Hilbert spaces of $2 \times N_v$
and $3 \times N_v$ dimensions.

\section{\label{sec:2timesN} Measures of entanglement between  electronic and vibrational degrees of freedom in a $2 \times N_v$ molecular system}

Considering two coupled electronic states ($g,e$), our aim is to derive an expression for the
electronic-nuclear entanglement produced by the vibronic coupling. We are especially interested
in describing cases of coupling due to a laser pulse, but the nature of the coupling does not
need to be specified in the formal part of our treatment. The only restriction that we employ
is that the coupling creates a pure state in the Hilbert space $\cal{H}$ $=$ $\cal{H}$$_{el}$
$\bigotimes$  $\cal{H}$$_{vib}$ of dimension $2 \times N_v$, whose wavefunction can be written as
\begin{equation}
  \Psi_{el,vib}(R,\vec{r_{i}};t) =
  \phi_{g}^{el}(\vec{r_{i}};R)\psi_{g}(R,t)
  + \phi_{e}^{el}(\vec{r_{i}};R)\psi_{e}(R,t).
\label{psielvib}
\end{equation}
If $\{|\chi_{v_g}(R)>\}_{v_g=\overline{1,N_g}}$ and $\{|\chi_{v_e}(R)>\}_{v_e=\overline{1,N_e}}$
are the orthonormal vibrational bases with dimensions $N_g$ and $N_e$ ($N_g+N_e=N_v$),
corresponding to the electronic surfaces $g,e$, respectively, Eq.~(\ref{psielvib}) can be rewritten
as
\begin{eqnarray}
|\Psi_{el,vib}(t) > = |g> \bigotimes \sum_{v_g=1}^{N_g} c_{v_g} (t) |\chi_{v_g}> \nonumber\\
+ |e> \bigotimes \sum_{v_e=1}^{N_e} c_{v_e} (t) |\chi_{v_e}>,
\label{pure-elvib}
\end{eqnarray}
where  $|g>$, $|e>$ designate the electronic states
$\phi_{g,e}^{el}(\vec{r_{i}};R)$, and the nuclear wavepackets $\psi_{g,e}(R,t)$ were developed in
their corresponding vibrational bases. The complex coefficients $\{c_{v_g} (t)\}, \{c_{v_e} (t)\}$
give the population probabilities $\{|c_{v_g} (t)|^2\}, \{|c_{v_e} (t)|^2\}$ of the vibrational
levels $\{v_g\}$ and $\{v_e\}$. For a closed system comprised of only these two electronic surfaces,
the normalization condition $ < \Psi_{el,vib}(t) |\Psi_{el,vib}(t) > =1$ is expressed by the relation
\begin{equation}
\sum_{v_g=1}^{N_g} |c_{v_g} (t)|^2 + \sum_{v_e=1}^{N_e} |c_{v_e} (t)|^2 =1,
\label{normvevg}
\end{equation}
and the density operator
\begin{equation}
 \hat{\rho}_{el,vib}(t)=|\Psi_{el,vib}(t) > <\Psi_{el,vib}(t)|
\label{densityop}
\end{equation}
corresponds to a pure state of the bipartite system: $\hat{\rho}^2_{el,vib}=\hat{\rho}_{el,vib}$.

Pure bipartite states have clear separability criteria like the Schmidt decomposition
\cite{4horodecki09, mintert-physrep05}, and  "good" measures of the amount of entanglement, the first
one being the von Neumann entropy of the reduced density matrix \cite{vedral02,vedralplenio98}.
Even if the von Neumann entropy of the subsystem is the "entropy of entanglement" \cite{bennett96}
for pure states, and it could be considered as "the unique measure for pure states" \cite{4horodecki09},
it was also argued that "only one measure is not sufficient to completely quantify entanglement of pure
states for bipartite systems", and "several independent measures should be employed simultaneously"
\cite{vidal00}. In the present work we shall refer to two measures quantifying the entanglement:
the von Neumann entropy and the linear entropy of the reduced density matrix.

To estimate the entanglement of $|\Psi_{el,vib}(t) >$, we have to analyze the reduced density operator
of one of the two subsystems: $\hat{\rho}_{el} =$ Tr$_{vib}(\hat{\rho}_{el,vib})$ or $\hat{\rho}_{vib}
=$ Tr$_{el}(\hat{\rho}_{el,vib})$. The spectrum of the reduced density matrix (for example $\hat{\rho}_{el}$)
gives the Schmidt coefficients  which allow to distinguish separable from entangled states and can be
used to obtain the von Neumann entropy $S_{vN}(\hat{\rho}_{el})$. On the other hand, the purity of the
reduced density,  Tr$\hat{\rho}^2_{el}$,  shows the degree of mixing of the subsystems and is also an
indicator for the degree of entanglement in system: if Tr$\hat{\rho}^2_{el}$ $\ne 1$ the state described
by Eq.~(\ref{pure-elvib}) is entangled \cite{qinfBarnett}.

In order to obtain a reduced density matrix one needs to designate an orthonormal basis set for each
subsystem Hilbert space, $\cal{H}$$_{el}$ and $\cal{H}$$_{vib}$. $ \{ |g> ,|e> \}$ constitutes such a
basis set for $\cal{H}$$_{el}$. In $\cal{H}$$_{vib}$ we have the two vibrational bases
$\{|\chi_{v_g}(R)>\}_{v_g=\overline{1,N_g}}$ and $\{|\chi_{v_e}(R)>\}_{v_e=\overline{1,N_e}}$, but
generally $<\chi_{v_g}(R)|\chi_{v_e}(R)> \ne 0$, so we need to construct a complete orthonormal vibronic
basis $ \{ |j > \}_{j=1,N_v}$ of  $\cal{H}$$_{vib}$, which will have the dimension $N_v=N_g+N_e$ and will
satisfy the orthonormality ($<j|j'>= \delta_{jj'}$) and completeness ($ \sum_{j=1}^{N_v} |j><j| = \hat{I}_v$)
conditions. Then, designating by $ \{ |1>,|2> \}$  a suited orthonormal basis in the electronic Hilbert
space $\cal{H}$$_{el}$, $|\Psi_{el,vib}(t) >$ can be also expressed as
\begin{eqnarray}
|\Psi_{el,vib}(t) > = |1> \bigotimes \sum_{j=1}^{N_v} C_{1j} (t) |j> \nonumber\\
+ |2> \bigotimes \sum_{j=1}^{N_v} C_{2j} (t) |j>.
\label{pure-elvib1}
\end{eqnarray}
The complex coefficients $C_{1j}, C_{2j}$ obey a normalization condition similar to Eq.~(\ref{normvevg}):
\begin{equation}
\sum_{j=1}^{N_v} (|C_{1j} (t)|^2 + |C_{2j} (t)|^2 )=1.
\label{normcj}
\end{equation}
The reduced density operator $\hat{\rho}_{el}$ can now be calculated using the vibronic basis
$\{ |j > \}_{j=1,N_v}$:
\begin{eqnarray}
\hat{\rho}_{el}(t)= \text{Tr}_{vib}(|\Psi_{el,vib}(t) > <\Psi_{el,vib}(t)|)\nonumber\\
=\sum_{j=1}^{N_v} <j|\Psi_{el,vib}(t) > <\Psi_{el,vib}(t)|j>,
\label{elreddens}
\end{eqnarray}
and the reduced density matrix $\left( \hat{\rho}_{el} \right)$ can be expressed in the electronic basis
$\{ |1>,|2> \}$ as
\begin{eqnarray}
&&\left( \hat{\rho}_{el} \right)=
\left(\begin{array}{lc}
  \sum_{j} |C_{1j}|^2 & \sum_{j} C_{1j} C^{*}_{2j} \\
  \sum_{j} C^{*}_{1j} C_{2j} & \sum_{j} |C_{2j}|^2
 \end{array} \right),
\label{matdensel}
\end{eqnarray}
where the summation is over $j=\overline{1,N_v}$, and Tr$\left( \hat{\rho}_{el} \right)=1$.

To express the quantities implying the coefficients $C_{1j}, C_{2j}$ as functions of entities related
to the initial electronic states $g,e$, we choose the new electronic basis set as
\begin{eqnarray}
|1> = \frac{1}{\sqrt{2}} ( |g> + |e>)  ~,~ |2> = \frac{1}{\sqrt{2}} (|g> - |e>).
\label{relbases}
\end{eqnarray}
Using Eqs.~(\ref{pure-elvib}), (\ref{pure-elvib1}), (\ref{relbases}) together with the orthornormality and
completeness relations, the quantities implying the coefficients $C_{1j}, C_{2j}$ can be expressed as
functions of $c_{v_g} (t)$,  $c_{v_e} (t)$ and of the vibrational eigenstates $|\chi_{v_g}>$, $|\chi_{v_e}>$.

The eigenvalues of the matrix (\ref{matdensel}) are $\rho_{+,-}(t)= \frac{1}{2} \{ 1 \pm [ P_g(t) - P_e(t)] \}$,
with $P_g(t)=\sum_{v_g} |c_{v_g} (t)|^2$ and $P_e(t)=\sum_{v_e} |c_{v_e} (t)|^2$ being the vibrational
populations of the $g,e$ electronic states. As $P_g(t)+P_e(t)=1$, the eigenvalues of the reduced density matrix
$(\hat{\rho}_{el})$ are simply the populations of the electronic states:
\begin{eqnarray}
\rho_{+}(t)=P_g(t) ~,~\rho_{-}(t)=P_e(t).
\label{eigenv}
\end{eqnarray}
Knowing that the eigenvalues $\rho_{+,-}(t)$ are the squares of the coefficients defining the Schmidt
decomposition of the pure bipartite state $|\Psi_{el,vib}(t)>$ \cite{4horodecki09,mintert-physrep05}, one reaches
the easily understandable conclusion that separability appears if only one of the electronic states is populated
($P_g(t) = 1$ and $P_e(t)=0$, or vice versa), and  maximum entanglement is realized for $P_g(t) = P_e(t) =1/2$.

To advance to the dynamical aspects of entanglement, one has to use measures such as the von Neumann entropy of
the reduced density matrix or the linear entropy (calculated via the purity of the reduced density matrix).
The von Neumann entropy of entanglement
\begin{equation}
S_{vN}(\hat{\rho}_{el}) = - \text{Tr} (\hat{\rho}_{el} \log_2 \hat{\rho}_{el})
\label{vonNentropy}
\end{equation}
is the Shannon entropy of the squares of the Schmidt coefficients \cite{bennett96}:
$S_{vN}(\hat{\rho}_{el}) = -\rho_{+} \log_2 \rho_{+} - \rho_{-} \log_2 \rho_{-}$.
Two alternative expressions can be written:
\begin{subequations}
\label{eq:vonNel}
\begin{equation}
S_{vN}(\hat{\rho}_{el}(t)) = - P_g(t) \log_2 P_g(t)  -  P_e(t)\log_2 P_e(t) \label{vonNel1}
\end{equation}
\begin{equation}
=  -\frac{1+D(t)}{2} \log_2 \frac{1+D(t)}{2} - \frac{1-D(t)}{2} \log_2  \frac{1-D(t)}{2}.
\label{vonNel2}
\end{equation}
\end{subequations}
The notation $D(t) = P_g(t) - P_e(t)$  was employed in deriving   (\ref{vonNel2}), which is the form
taken by the von Neumann entropy for the density operator of a qubit, $D(t)$ being the module of the Bloch
vector \cite{qinfBarnett}.

As expected, the von Neumann entropy $S_{vN}(\hat{\rho}_{el}(t))$ is 0 for a separable state (if one of the
eigenvalues is 1, the other being 0), and attains the maximum value 1 for maximum entanglement (when $P_g(t)
 = P_e(t) =1/2$).  It is important to notice that  Eq.~(\ref{vonNel1}) gives the possibility to
investigate the entanglement dynamics in a molecular process.

Now we shall analyze the purity of the reduced density matrix, Tr$(\hat{\rho}^2_{el}(t))$, which is related
to the linear entropy of entanglement $L(t)$:
\begin{equation}
L(t)=1-\text{Tr}(\hat{\rho}^2_{el}(t)).
\label{linentropy}
\end{equation}
Using Eq.~(\ref{matdensel}) one obtains
\begin{equation}
\text{Tr} (\hat{\rho}^2_{el}) = (\sum_{j=1}^{N_v} |C_{1j} |^2 )^2 + (\sum_{j=1}^{N_v} |C_{2j} |^2 )^2 +
 2 | \sum_{j=1}^{N_v} C_{1j} C^{*}_{2j} |^2.
\label{tracero2}
\end{equation}
These quantities can be written as functions of $c_{v_g}(t)$,  $c_{v_e}(t)$, $|\chi_{v_g}>$, $|\chi_{v_e}>$,
to reach the expression
\begin{eqnarray}
\text{Tr} (\hat{\rho}^2_{el}(t)) = \frac{1}{2} + \frac{1}{2} [P_g(t) - P_e(t)]^2 \nonumber\\
+2 | \sum_{v_g=1}^{N_g}\sum_{v_e=1}^{N_e} c^{*}_{v_g} (t) c_{v_e} (t)  <\chi_{v_g} (R)|\chi_{v_e}(R)> |^2.
\label{purityred1}
\end{eqnarray}
Using the condition $P_g(t)+P_e(t)=1$, and writing the vibrational wavepacket in an electronic state as
\begin{eqnarray}
 |\psi(R,t)>= \sum_{v} c_{v} (t) |\chi_{v}(R)>,
\end{eqnarray}
Eq.~(\ref{purityred1}) takes the simple form
\begin{equation}
\text{Tr} (\hat{\rho}^2_{el}(t)) = P^2_g(t) + P^2_e(t) + 2 |<\psi_{g}(R,t)|\psi_{e}(R,t)>|^2.
\label{purityred2}
\end{equation}
Eqs.~(\ref{purityred1}), (\ref{purityred2}) show the purity Tr$(\hat{\rho}^2_{el}(t))$ as an interesting
sensor for the correlations between the electronic channels, emphasizing explicitly the role played by
the vibronic coherences. The purity defined by Eq.~(\ref{purityred1}) is bounded by $\frac{1}{2} \le$
Tr$(\hat{\rho}^2_{el}(t)) \le 1$
\footnote{Generally, the purity Tr$\hat{\rho}^2$  of a quantum state is bounded by $\frac{1}{d} \le$
Tr$\hat{\rho}^2\le 1$, where $d$ is the dimension of the Hilbert space attributed to the system
\cite{bookJaeger}.}
and the corresponding linear entropy by $0 \le L(t) \le \frac{1}{2}$. If Tr$(\hat{\rho}^2_{el}(t)) =1$
(and $L(t)=0$) the electronic subsystem is pure by itself, and then the pure bipartite state is non-entangled.
It is, obviously, the result obtained with the relation (\ref{purityred1}) if only one of the electronic
states is populated (all $c_{v_g}(t)=0$ or all $c_{v_e}(t)=0$).

These results show that an interaction between two electronic channels which leaves both channels populated
will produce an entangled state, entanglement being present at all times if both channels remain populated.

In the following, we will use the expressions obtained for $S_{vN}(\hat{\rho}_{el}(t))$, Tr$(\hat{\rho}^2_{el}(t))$,
and $L(t)$
\footnote{Comparing entanglement measures is obviously a subtle and complicated matter. We just
make the observation that the von Neumann entropy $S_{vN}(\hat{\rho})$ and the purity Tr$(\hat{\rho}^2)$
are both related to the quantum $\alpha$ Renyi entropies $S_{\alpha}(\hat{\rho})=(1-\alpha)^{-1} \log_2$Tr$
\hat{\rho}^{\alpha}$. In the limit $\alpha \to 1$, one obtains the entropy of entanglement: $S_1(\hat{\rho})
=S_{vN}(\hat{\rho})$. On the other hand, $S_2(\hat{\rho})=- \log_2$ Tr$(\hat{\rho}^2)$. Nevertheless, the
Renyi entropies for $\alpha > 1$ do not have the same mathematical properties as those with $0 \le \alpha
\le 1$, which fulfill the maximum of postulates required for an entanglement measure \cite{vidal00, 4horodecki09}.}
to analyze the entanglement dynamics in specific cases of coupling of two electronic channels by a laser pulse.

\subsection{\label{sec:2times2} Entanglement dynamics produced by a constant vibronic coupling in a $2 \times 2$ system:
one vibrational level in each electronic state}

We consider first the model case of two electronic states $g,e$ coupled by an electric field with amplitude
${\cal{E}}(t)={\cal {E}}_0 \cos \omega_Lt$.  In the rotating wave approximation, the evolution of such a
system is described by a time-dependent Schr\"odinger equation like Eq.~(\ref{tschreq2}), but with constant
coupling $W_L$ \cite{vatasescu01}. Considering a $2 \times 2$ system with one vibrational state associated
with every electronic state, the "vibrational wavepackets" associated with the electronic states are
$|\psi_{g}(R,t)>= c_{v_g}(t)|\chi_{v_g} (R)>$ and $|\psi_{e}(R,t)> = c_{v_e}(t)|\chi_{v_e} (R)>$ (with
$|c_{v_g}(t)|^2 + |c_{v_e}(t)|^2 =1$). In this case one can write an analytic expression for the population
$|c_{v_e}(t)|^2$, showing the Rabi beats induced by the coupling between the two vibrational states with
energies $E_{v_e}$, $E_{v_g}$ \cite{vatasescu09}:
\begin{eqnarray}
|c_{v_e}(t)|^2= \frac{|W_L \mathcal{F}_{v_g v_e} |^2}{(\hbar \Omega_{v_e,v_g})^2}
\sin^2(\Omega_{v_e,v_g} t), \label{cve2}
\\
\hbar \Omega_{v_e,v_g} = \sqrt{ |W_L \mathcal{F}_{v_g v_e}|^2 + [(E_{v_e}-E_{v_g})/2]^2 }.
\label{Omegavevg}
\end{eqnarray}
In Eqs.~(\ref{cve2}), (\ref{Omegavevg}) $|\mathcal{F}_{v_g v_e}|^2$ is the Franck-Condon factor,
with  $\mathcal{F}_{v_g v_e}=<\chi_{v_g} (R)|\chi_{v_e}(R)>$ the overlap integral of the vibrational wavefunctions.

The eigenvalues of the reduced density matrix $\hat{\rho}_{el}(t)$ are the populations of the two
electronic states: $\rho_{+}(t)=P_g(t)=|c_{v_g}(t)|^2$,  $\rho_{-}(t)=P_e(t)=|c_{v_e}(t)|^2$.
Then, according to Eqs.~(\ref{eq:vonNel}), (\ref{purityred1}), the von Neumann entropy and the
purity are:
\begin{equation}
S_{vN}(\hat{\rho}_{el}(t)) = -|c_{v_g}(t)|^2 \log_2 |c_{v_g}(t)|^2  - |c_{v_e}(t)|^2 \log_2 |c_{v_e}(t)|^2,
\label{SvN22}
\end{equation}
\begin{equation}
\text{Tr}(\hat{\rho}^2_{el}(t)) = 1-2(1-|\mathcal{F}_{v_g v_e}|^2) |c_{v_g}(t)|^2 |c_{v_e}(t)|^2.
\label{purity22}
\end{equation}
The linear entropy of entanglement becomes
\begin{equation}
L_{v_g v_e}(t) = 2(1-|\mathcal{F}_{v_g v_e}|^2) |c_{v_e}(t)|^2 (1-|c_{v_e}(t)|^2),
\end{equation}
with $|c_{v_e}(t)|^2$ given by Eq.~(\ref{cve2}), which means that the characteristic period appearing
in the linear entropy evolution is the Rabi period $T^R_{v_e,v_g}$ of the beating between the vibrational
levels ($v_e$,$v_g$), shown by Eq.~(\ref{cve2}):
\begin{equation}
T^R_{v_e,v_g}=\frac{\pi}{\Omega_{v_e,v_g}}.
\label{vRabi}
\end{equation}
By showing that the Rabi period associated to the vibronic coupling is the characteristic time in the
evolution of the linear entropy, already this simple model  provides insight into the entanglement dynamics
during the laser coupling. A beat phenomenon in the reduced-density linear entropy is also signalled in
Ref.~\cite{hou06}, which analyzes the entanglement of vibrations in triatomic molecules using an algebraic model.

\subsection{\label{sec:dyn1g} Entanglement dynamics in the case of two electronic states coupled by a laser pulse.
Example of $a^3\Sigma_{u}^{+}(6s,6s)=g$, $1_g(6s,6p_{3/2})=e$ of the Cs$_2$ molecule}

\begin{figure}
\includegraphics[width=0.95\columnwidth]{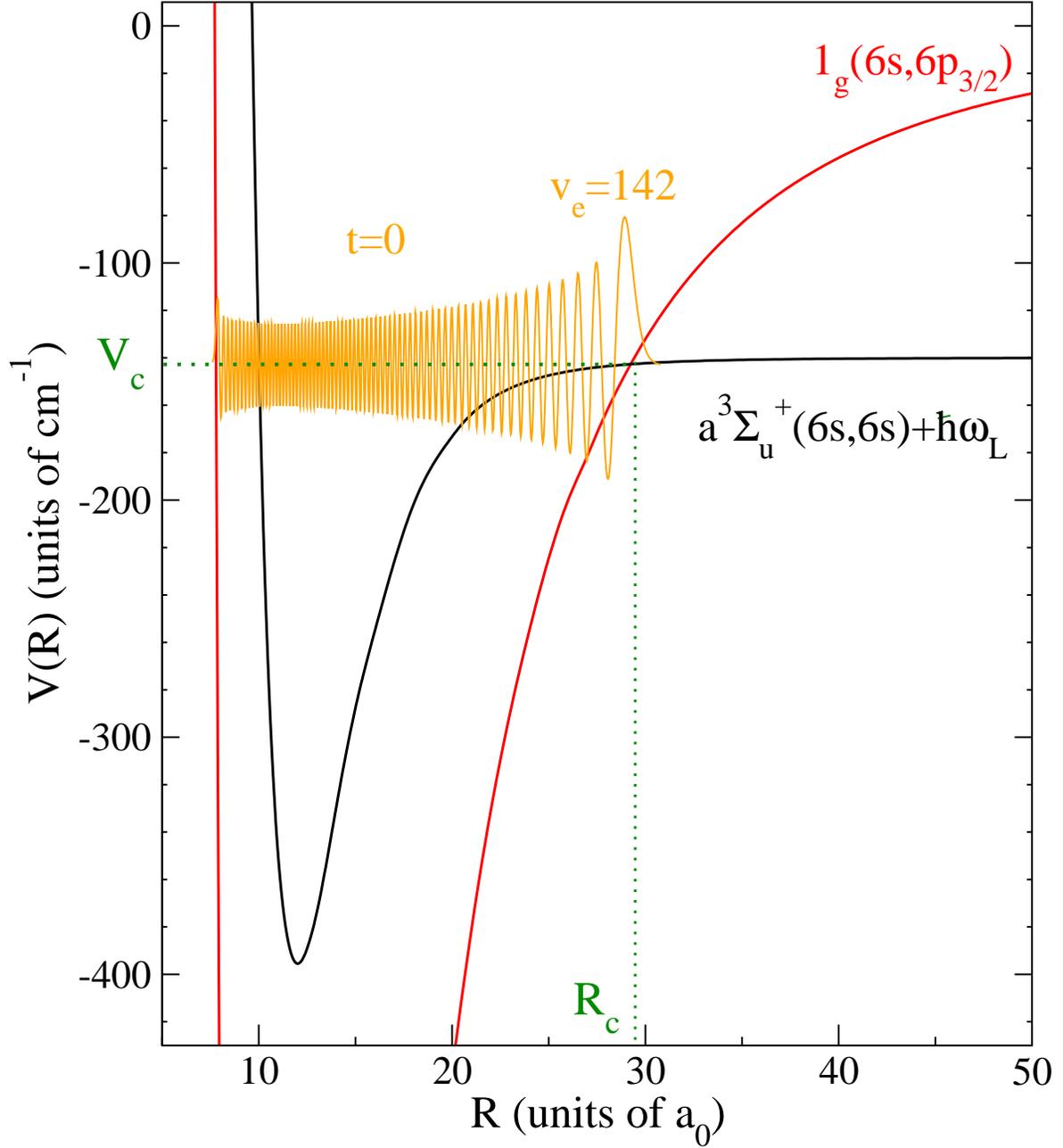}%
 \caption{\label{fig1-potS1gwf142} (Color online) $a^3\Sigma_{u}^{+}(6s,6s)$ and $1_g(6s,6p_{3/2})$ electronic
potentials of Cs$_2$, dressed with the photon energy $\hbar\omega_L$= $E_{6p_{3/2}}-E_{6s} - \hbar\Delta_L$
($\hbar \Delta_L$=140 cm$^{-1}$) and crossing at $R_c=29.3 \ a_0$, $V_c$=$V_{1_g}(R_c)$=$V_{\Sigma}(R_c)$=-143
cm$^{-1}$.  The energy origin is taken to be the dissociation limit $E_{6s+6p_{3/2}}=0$ of the $1_g(6s+6p_{3/2})$
potential. The initial wavefunction of the process is the vibrational wavefunction with $v_e=142$ of the $1_g$
electronic state, also represented in the figure.}
\end{figure}

\begin{figure}
\includegraphics[width=0.95\columnwidth]{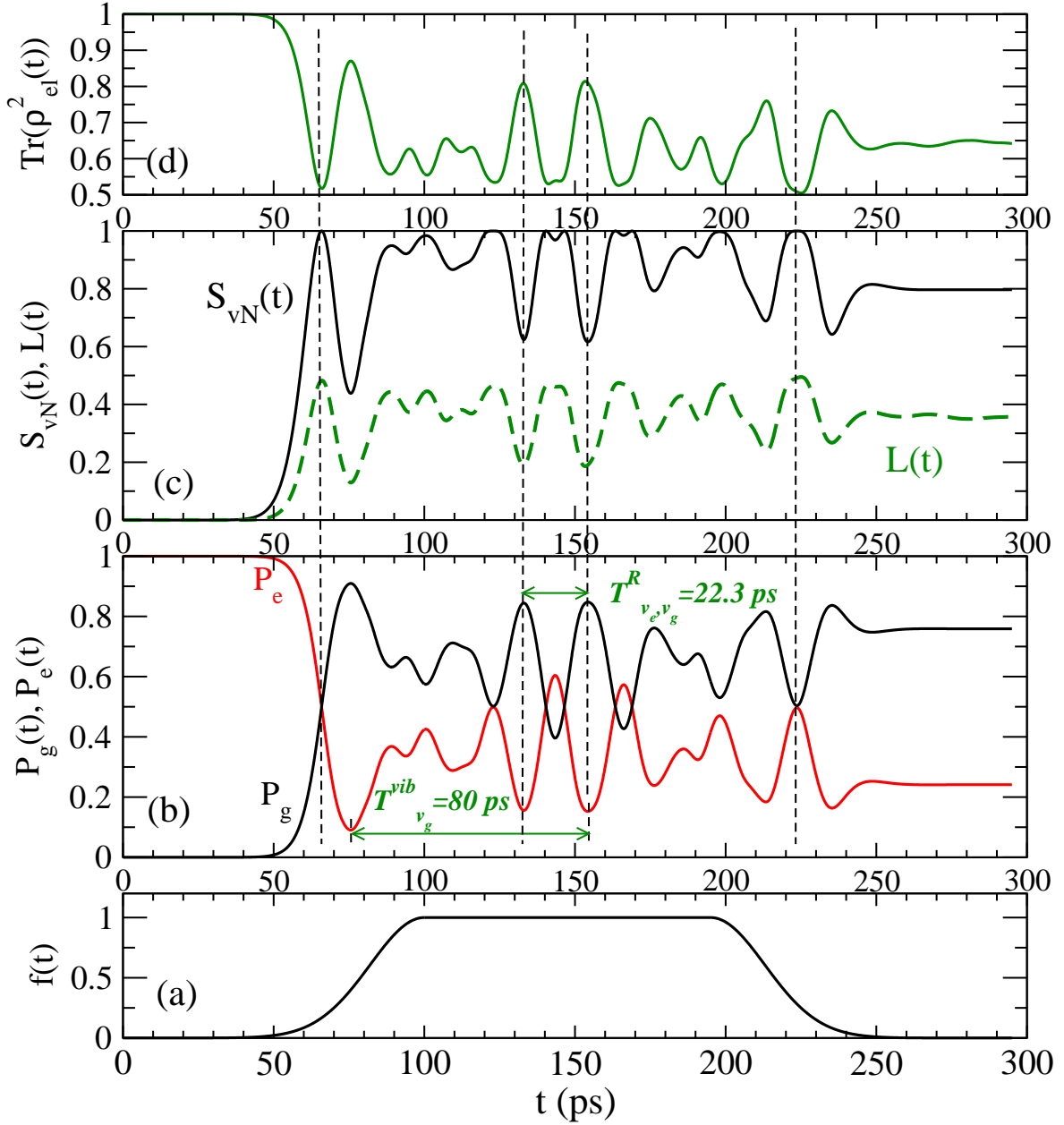}%
 \caption{\label{fig2-ent2PotS1g} (Color online) Dynamics of entanglement between electronic and vibrational
degrees of freedom for the electronic states $g=a^3\Sigma_{u}^{+}$ and $e=1_g$ of Cs$_2$ (Fig.~\ref{fig1-potS1gwf142})
coupled by a laser pulse. (a) Pulse envelope $f(t)$. (b) Time evolution of the populations $P_g(t)$ and $P_e(t)$. (c)
Time evolution of the von Neumann entropy $S_{vN}(t)$ (full line) and linear entropy $L(t)$ (dashed line). (d)
Evolution of the purity $Tr (\hat{\rho}^2_{el}(t))$.}
\end{figure}

Here we consider the intramolecular dynamics induced by a laser pulse which couples two electronic
states of the Cs$_2$ molecule. This allows one to follow the temporal dependence of the entanglement
measures proposed in the preceding sections and to relate it to the characteristic times of the
molecular evolution.

We consider that the electronic channels $e=1_g(6s,6p_{3/2})$ and $g=a^3\Sigma_u^+(6s,6s)$ are coupled
by an electric field with amplitude ${\cal{E}}(t)={\cal {E}}_0 f(t) \cos \omega_Lt$, such that the potential
curves dressed with the photon energy $\hbar\omega_L$ have a crossing point at $R_c$ (Fig.~\ref{fig1-potS1gwf142}).
The time-dependent Schr\"odinger equation associated with the radial motion of the wavepackets $\Psi_{e}(R,t)$
and $\Psi_{g}(R,t)$ in the electronic channels, written using the rotating wave approximation with the frequency
$\omega_L/2\pi$ \cite{vatasescu09,vatasescu01}, is:
\begin{eqnarray}
\label{tschreq2}
&&i\hbar\frac{\partial}{\partial t}\left(\begin{array}{c}
 \Psi_{e}(R,t)\\
\Psi_{g}(R,t)
 \end{array}\right)=\\
&&
\left(\begin{array}{lc}
 {\bf \hat T} + V_{e}(R)  &
W_L f(t) \\
 W_L f(t) &
 {\bf \hat T} + V_{g}(R)
 \end{array} \right)
 \left( \begin{array}{c}
 \Psi_e(R,t)\\
\Psi_g(R,t)
 \end{array} \right). \nonumber
 \end{eqnarray}
The potentials $V_{e}(R)$ and $V_{g}(R)$ are the diabatic electronic potentials crossing at $R_c$,
represented in Fig.~\ref{fig1-potS1gwf142}. ${\bf \hat T}$ is the kinetic energy operator and $W_L f(t)$
the coupling between the two channels, with $f(t)$ the temporal envelope of the pulse (shown in Fig.~
\ref{fig2-ent2PotS1g}(a)). $W_L= - \frac {1}{2}{\cal {E}}_0 D_{ge}^{\vec{e_L}}$, where ${\cal {E}}_0=
\sqrt{2I / c\epsilon_0}$ is the field amplitude (with $I$ the laser intensity), ${\vec{e_L}}$ the
polarization, and $D_{ge}^{\vec{e_L}}$  the transition dipole moment between the ground and the excited
molecular electronic states. If one neglects the R-dependence of the transition dipole moment, using a
value $D_{ge}^{\vec{e_L}}$ deduced from standard long-range calculations for a linear polarisation vector
$\vec{e_L}$  \cite{vatasescu99}, for a pulse intensity I $\approx$ 43 MW/cm$^{2}$ one obtains a coupling
W$_L$=13.17 cm$^{-1}$. The initial state of the process, represented in Fig.~\ref{fig1-potS1gwf142}, is
the vibrational eigenstate $\chi^{v_e=142}_{1_g}(R)$ corresponding to the vibrational level $v_e=142$ of
the excited electronic potential $1_g(6s+6p_{3/2})$,  bound by $E_{v_e}=-140.9$ cm$^{-1}$.

The dynamics is simulated solving numerically the Schr\"odinger equation (\ref{tschreq2}), by propagating
in time the initial wavefunction
$\left(\begin{array}{c}
\chi^{v_e=142}_{1_g}(R) \\
0
\end{array}\right)$
on a spatial grid  with length $L_R \approx 370$ a$_0$. The time propagation uses the Chebychev expansion
of the evolution operator  \cite{kosloff94,kosloff96} and the Mapped Sine Grid (MSG) method \cite{elianeepjd04,
 willner04}  to represent the radial dependence of the wavepackets. The populations in each electronic state
are calculated from the vibrational wavepackets $\Psi_{e}(R,t)$ and $\Psi_{g}(R,t)$ as $P_{g,e}(t) = \int^{L_R}
| \Psi_{g,e}(R',t) |^{2} dR'$, with the total population normalized at 1 on the spatial grid ($P_g(t)+P_e(t)=1$),
and $P_e(0)=1$.

The evolution of the populations $P_g(t)$, $P_e(t)$ during the pulse is shown in Fig.~\ref{fig2-ent2PotS1g}(b).
The chosen pulse is sufficiently long (about 200 ps) and strong (the maximum local Rabi period associated with
the constant coupling $W_L$ is $T_{Rabi}(R_c)=\hbar \pi / W_L$=1.27 ps \cite{vatasescu09}) to put in evidence
typical phenomena such as the beats between various vibrational levels populated by the pulse, and the vibrational
motion in the potential wells. Several vibrational levels of each electronic surface having energies close to the
energy crossing $V_c$  are populated during the pulse, with typical vibrational periods of 11 ps in the $1_g$
potential, and between 40 ps and 80 ps in the $a^3\Sigma_{u}^{+}$ potential. The timescales related to the laser
coupling and vibrational motion have been analyzed in detail in Ref.~\cite{vatasescu09}. The time evolution shown in Fig.~\ref{fig2-ent2PotS1g}(b) is characterized by inversion of population between the two channels and a Rabi beating
with the period $T^R_{v_e,v_g}=22.3$ ps, specific to the vibrational levels $v_e=142$ of $1_g$ and $v_g=47$ of
$a^3\Sigma_{u}^{+}$, whose vibrational periods are $T^{vib}_{v_e=142}=10.8$ ps and $T^{vib}_{v_g=47}=80$ ps.
In the figure, this last characteristic time appears as related to the revival of Rabi oscillations of maximum amplitude.

The entanglement dynamics is illustrated by the evolution of the von Neumann entropy $S_{vN}(t)$ and the linear
entropy $L(t)$, represented in Fig.~\ref{fig2-ent2PotS1g}(c). Both show similar time oscillations, with periods
which are those of the beats between the populations $P_g(t)$ and $P_e(t)$, dominated here by the Rabi period
$T^R_{v_e,v_g}=22.3$ ps. The equalization of populations between the two electronic channels creates the condition
for maximum entanglement ($S_{vN}(t)=1$), which is repeatedly realized during the pulse action. Finally, the laser
pulse leaves the system in an entangled state $|\Psi_{el,vib}(t) >$ characterized by a high von Neumann entropy
$S_{vN}(t=300$ ps)$\approx 0.8$.

\section{\label{sec:3electr}Entanglement between electronic and vibrational degrees of freedom in a $3 \times N_v$ system.}

We consider now a bipartite Hilbert space $\cal{H}=$$\cal{H}$$_{el}$ $\bigotimes$ $\cal{H}$$_{vib}$ of
dimension $3 \times N_v$. We assume the existence of couplings between the three electronic states
$|g>$, $|e>$, $|f>$, such that a pure state of $\cal{H}$ is created:
\begin{eqnarray}
|\Psi_{el,vib}(t) > = |g> \bigotimes \sum_{v_g=1}^{N_g} c_{v_g}(t)|\chi_{v_g}>\nonumber\\
+ |e> \bigotimes \sum_{v_e=1}^{N_e} c_{v_e} (t) |\chi_{v_e}> \nonumber\\
 + |f> \bigotimes \sum_{v_f=1}^{N_f} c_{v_f} (t) |\chi_{v_f}>.
\label{pure-elvib3}
\end{eqnarray}
$\{|\chi_{v_g}>\}_{v_g=\overline{1,N_g}}$, $\{|\chi_{v_e}>\}_{v_e=\overline{1,N_e}}$, and
$\{|\chi_{v_f}>\}_{v_f=\overline{1,N_f}}$  are the orthonormal vibrational bases (with dimensions $N_g$,
$N_e$, $N_f$, respectively) corresponding to the electronic surfaces $g,e,f$, and the dimension of the
vibrational Hilbert space is $N_v=N_g+N_e+N_f$. The normalization condition  $<\Psi_{el,vib}(t)|\Psi_{el,
vib}(t)>=1$ is expressed by the relation
\begin{equation}
\sum_{v_g} |c_{v_g} (t)|^2 + \sum_{v_e} |c_{v_e} (t)|^2 + \sum_{v_f} |c_{v_f} (t)|^2=1,
\label{normvevg3}
\end{equation}
and the density operator $\hat{\rho}_{el,vib}(t)$ associated with the pure state (as in Eq.~(\ref{densityop}))
obeys $\hat{\rho}^2_{el,vib}=\hat{\rho}_{el,vib}$. Following the line of reasoning employed in Sec.~\ref{sec:2timesN},
the quantification of the entanglement requires the calculation of the electronic reduced density matrix,
for which we have to use a complete orthonormal vibronic basis $ \{ |j > \}_{j=1,N_v}$, satisfying the
orthonormality ($<j|j'>= \delta_{jj'}$) and completeness ($ \sum_{j=1}^{N_v} |j><j| = \hat{I}_v$) conditions
in  $\cal{H}$$_{vib}$. This vibronic basis can be associated to a new orthonormal electronic basis $ \{ |1>,|2>,
|3> \}$ in $\cal{H}$$_{el}$, such that the wavefunction $|\Psi_{el,vib}(t)>$ of the pure bipartite system may be
also expressed as
\begin{eqnarray}
|\Psi_{el,vib}(t) > = |1> \bigotimes \sum_{j=1}^{N_v} C_{1j} (t) |j>\nonumber\\
+ |2> \bigotimes \sum_{j=1}^{N_v} C_{2j} (t) |j> \nonumber\\
+ |3> \bigotimes \sum_{j=1}^{N_v} C_{3j} (t) |j>.
\label{pure-elvib3j}
\end{eqnarray}
The complex coefficients $C_{1j}, C_{2j}, C_{3j}$ obey the normalization condition $\sum_{j=1}^{N_v}
(|C_{1j} (t)|^2 + |C_{2j} (t)|^2  + |C_{3j} (t)|^2 )=1$.

The reduced density operator $\hat{\rho}_{el}$ is calculated using the vibronic basis $\{ |j > \}_{j=1,N_v}$
as shown in Eq.~(\ref{elreddens}), and the reduced density matrix $\left( \hat{\rho}_{el} \right)$ can be
written in the electronic basis $ \{ |1>,|2>, |3>\}$:
\begin{eqnarray}
&&\left( \hat{\rho}_{el} \right)=
\left(\begin{array}{ccc}
\sum_{j} |C_{1j}|^2 & \sum_{j} C_{1j} C^{*}_{2j} &  \sum_{j} C_{1j} C^{*}_{3j} \\
\sum_{j} C_{2j} C^{*}_{1j}  & \sum_{j} |C_{2j}|^2 &  \sum_{j} C_{2j} C^{*}_{3j} \\
\sum_{j} C_{3j} C^{*}_{1j}  &  \sum_{j} C_{3j} C^{*}_{2j}  &  \sum_{j} |C_{3j}|^2
\end{array} \right),
\label{matdensel3}
\end{eqnarray}
where the j-summations are  over $j=\overline{1,N_v}$. It is nevertheless difficult to diagonalize the
reduced density matrix (\ref{matdensel3}) and to obtain its eigenvalues in order to get an analytic expression
for the von Neumann entropy $S_{vN}(t)$. However, the purity of the reduced density matrix can be calculated
using Eq.~(\ref{matdensel3}):
\begin{eqnarray}
\text{Tr}(\hat{\rho}^2_{el}) = (\sum_{j} |C_{1j} |^2 )^2 + (\sum_{j} |C_{2j} |^2 )^2 +
(\sum_{j} |C_{3j} |^2 )^2 \nonumber\\
+ 2 \left( | \sum_{j} C_{1j} C^{*}_{2j} |^2
+ | \sum_{j} C_{1j} C^{*}_{3j} |^2 +
| \sum_{j} C_{2j} C^{*}_{3j} |^2 \right). \nonumber\\
\label{tracero3j}
\end{eqnarray}

Making a choice for the new electronic basis set, as for example:
\begin{eqnarray}
|1> = \frac{1}{\sqrt{3}} (|g> + |e> + |f>),\nonumber\\
|2> = \frac{1}{\sqrt{6}} (|g> + |e> - 2|f>),\nonumber\\
|3> = \frac{1}{\sqrt{2}} (|g> - |e>),
\label{relbases3}
\end{eqnarray}
allows one to arrive in Eq.~(\ref{tracero3j}) at an expression related to the initial electronic states,
based on the electronic populations $P_g(t), P_e(t), P_f(t)$ and overlaps between $|\psi_{g,e,f}(R,t)>$:
\begin{eqnarray}
\text{Tr}(\hat{\rho}^2_{el}(t)) = P^2_g(t)+ P^2_e(t)+ P^2_f(t)\nonumber \\
+ 2 |<\psi_{g}(R,t)|\psi_{e}(R,t)>|^2 \nonumber \\
+ 2|<\psi_{g}(R,t)|\psi_{f}(R,t)>|^2  \nonumber \\
+ 2|<\psi_{e}(R,t)|\psi_{f}(R,t)>|^2.
\label{purityred3}
\end{eqnarray}

Eq.~(\ref{purityred3}) has the same structure as Eq.~(\ref{purityred2}), which now can be regarded as its
particular case for only two populated electronic states. The purity given by Eq.~(\ref{purityred3}) is
bounded by $\frac{1}{3} \le$ Tr$(\hat{\rho}^2_{el})\le 1$, which gives boundaries $0 \le L(t) \le \frac
{2}{3}$ for the linear entropy. We shall use Eq.~(\ref{purityred3}) to quantify the electronic-vibrational
entanglement in a $3 \times N_v$ molecular system using the reduced linear entropy, $L(t)=1-\text
{Tr}(\hat{\rho}^2_{el}(t))$. The next section constitutes an example.

\section{\label{sec:3elchirp} Entanglement dynamics in a system of three electronic states coupled by a sequence of two chirped laser pulses}

\begin{figure}
\includegraphics[width=0.9\columnwidth]{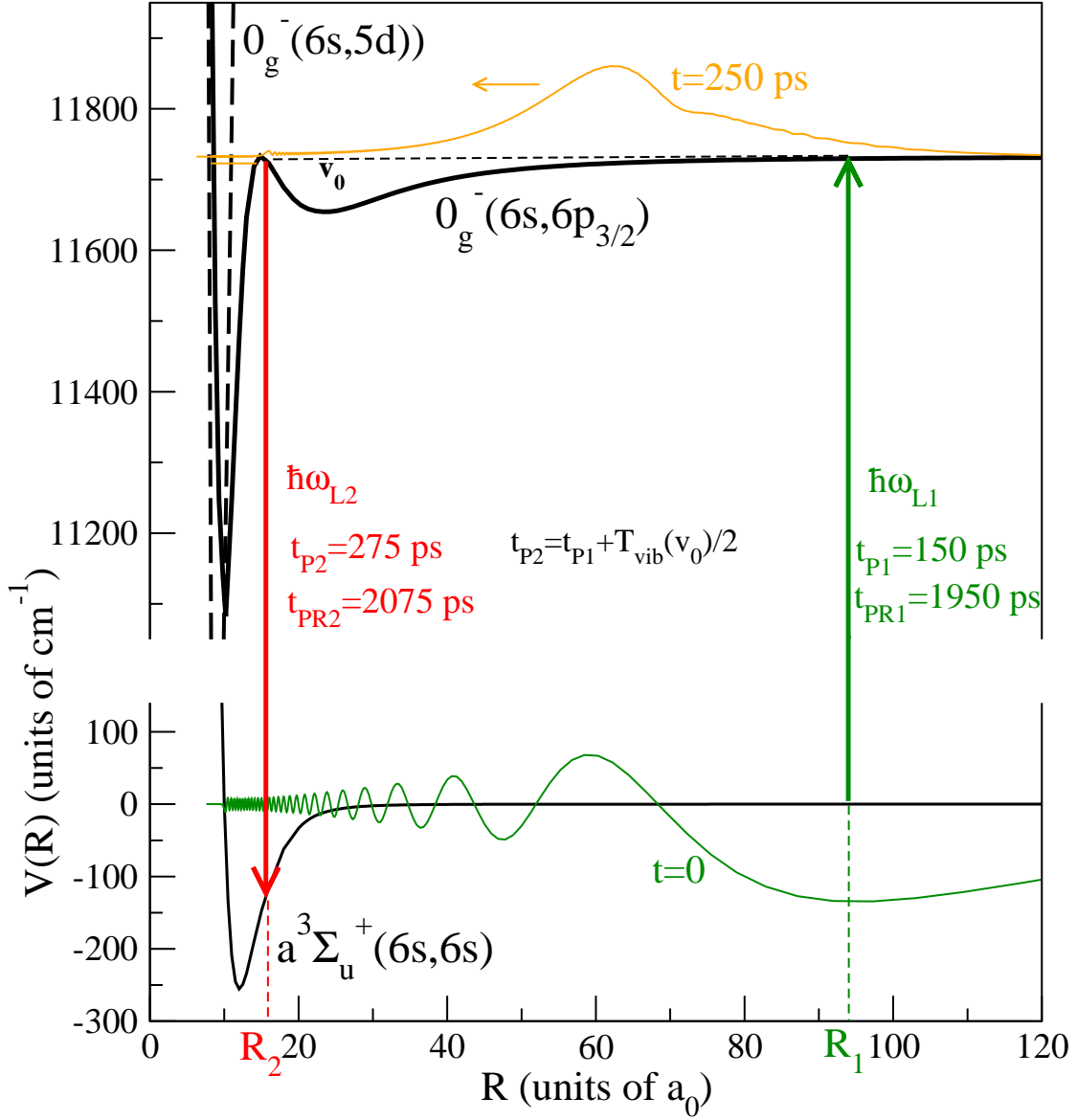}%
\caption{\label{fig3-pot3su0g6p5d}(Color online) Electronic potentials $a^3\Sigma_{u}^{+}(6s,6s)=g$, $0_g^-(6s,6p_{3/2})=e$,
and $0_g^-(6s,5d)=f$ of the Cs$_2$ molecule, coupled by a sequence of two chirped laser pulses (see also
Fig.~\ref{fig4-pulsesrep}) with central frequencies $\omega_{L1}/2\pi$ and $\omega_{L2}/2\pi$, and by a non-adiabatic
coupling (to be seen in the crossing of $e$ and $f$ electronic surfaces at small interatomic distances of about 10 $a_0$).
The initial state at t=0 (a loosely bound vibrational state in $g=a^3\Sigma_{u}^{+}$) is also shown, as well as the
vibrational wavepacket created by the first pulse in the $e$ state, at t=250 ps.}
\end{figure}

In this section we analyze the production of entanglement (quantified by the linear entropy $L(t)$)
in the case of a more complex molecular dynamics, related to a theoretical control scheme proposed
to create Cs$_2$ vibrationally cold molecules \cite{chirpPRA04,vatasescu06,vatasescu12} using the
multichannel tunneling observed in the cesium photoassociation spectrum \cite{vatasescu00}.

\begin{figure}
\includegraphics[width=0.9\columnwidth]{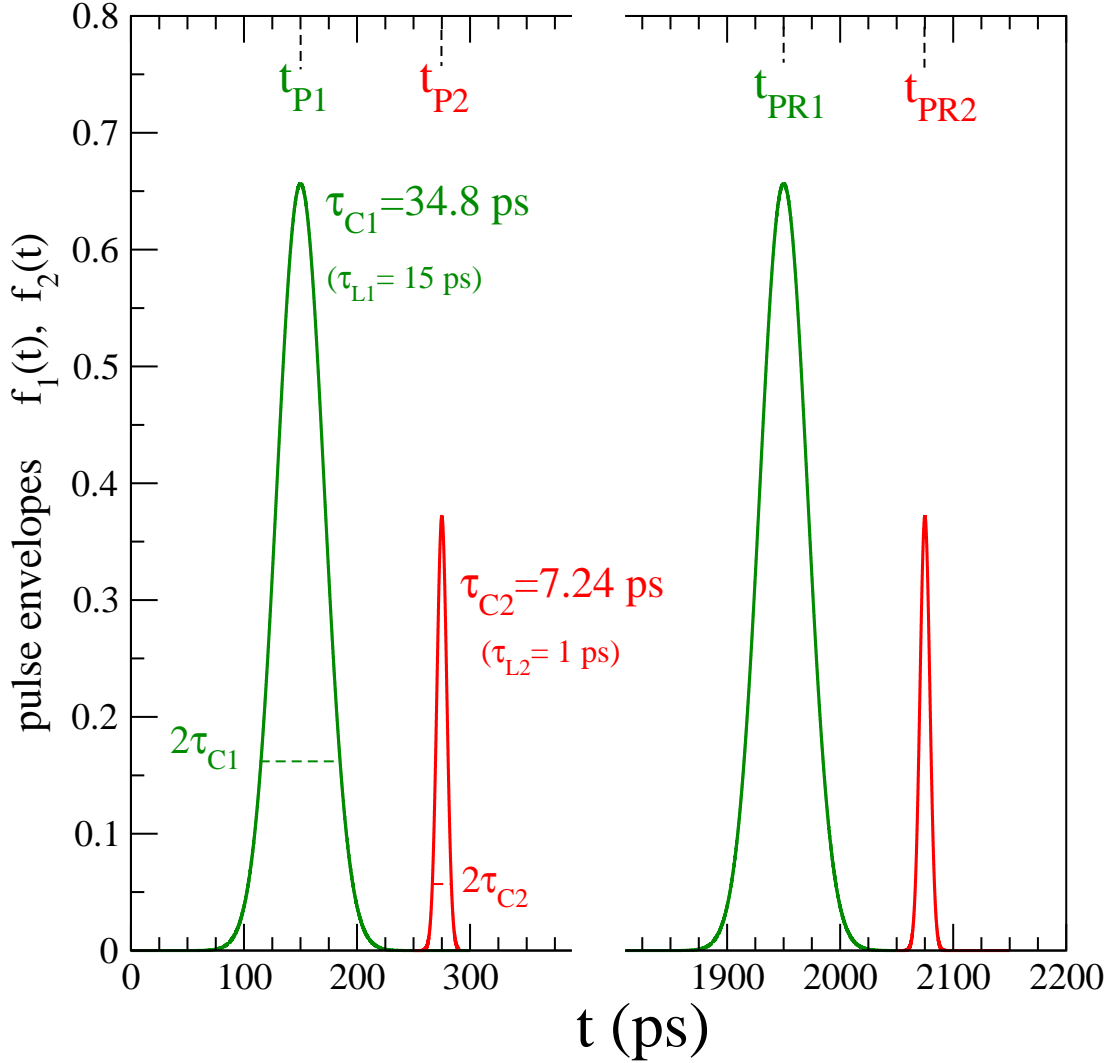}%
\caption{\label{fig4-pulsesrep} (Color online) The Gaussian temporal envelopes $f_1(t)$, $f_2(t)$ of the
chirped laser pulses which couple the electronic potentials in Fig.~\ref{fig3-pot3su0g6p5d}.
The maximum of an envelope is at $f(t_P)=\sqrt{\tau_L/ \tau_C}$.
The first sequence is made of two pulses centered in $t_{P1}=150$ ps and $t_{P2}=275$ ps.
The repetition of the sequence after 1800 ps is also shown ($t_{PR1}=1950$ ps and $t_{PR2}=2075$ ps.)}
\end{figure}

The scheme, illustrated in Fig.~\ref{fig3-pot3su0g6p5d}, uses a sequence of two chirped laser pulses
to couple the electronic potentials $a^3\Sigma_{u}^{+}(6s,6s)=g$ and $0_g^-(6s,6p_{3/2})=e$ of a Cs$_2$
cold molecule at large interatomic distances ($R_1 \approx 94 \ a_0$), as well as at small distances
($R_2 \approx 15.6 \ a_0$), in order to capture vibrational population in low vibrational levels of the
electronic potentials. Moreover, the $0_g^-(6s,6p_{3/2})=e$ state (having a double-well potential)
is coupled in the inner well region to the $0_g^-(6s,5d)=f$ electronic state, through a non-adiabatic
coupling generated by the spin-orbit interaction \cite{vatasescu06}. The first chirped pulse, with central
frequency $\omega_{L1}/2\pi$ ($\hbar \omega_{L1}=11729.66$ cm$^{-1}$) at $t_{P1}=150$ ps, couples $a^3
\Sigma_{u}^{+}(6s,6s)$ and $0_g^-(6s,6p_{3/2})$ at large interatomic distances ($R_1 \approx 94 \ a_0$).
We consider as initial state of the process the "last bound state" of the  $a^3\Sigma_{u}^{+}(6s,6s)$
potential obtained on a spatial grid of about 1060 $a_0$. Its wavefunction (partially visible in
Fig.~\ref{fig3-pot3su0g6p5d}) extends up to about 350 $a_0$ and has a maximum at $R_1$, being an advantageous
choice for the simulation of a cold photoassociation process. Operating on this initial state, the first
pulse creates a vibrational wavepacket around the vibrational level $v_0=98$ of the $0_g^-(6s,6p_{3/2})$
outer well \cite{chirpPRA04}, which begins to move to small distances  in the $0_g^-(6s+6p_{3/2})$ double
well (see Fig.~\ref{fig3-pot3su0g6p5d}). The second delayed pulse, with $\hbar \omega_{L2}=11856.66$ cm
$^{-1}$ and $t_{P2}=275$ ps \cite{vatasescu12}, induces a coupling in the zone of the $0_g^-(6s,6p_{3/2})$
double well barrier ($R_2 \approx 15.6 \ a_0$), controlling the tunneling in the $0_g^-(6s,6p_{3/2})$
potential coupled radially at small interatomic distances ($\approx 10 \ a_0$) with the $0_g^-(6s,5d)$
potential, and transferring population in low vibrational levels of the $a^3\Sigma_{u}^{+}(6s,6s)$ state.
The theoretical model and the choice of the chirped pulses are described in detail in Refs.~\cite{chirpPRA04,
vatasescu12}. Our present calculations include in the model the non-adiabatic coupling at short distances
between the $0_g^-(6s,6p_{3/2})$ and $0_g^-(6s,5d)$ potentials, and the repetition of the pulses sequence
after 1800 ps (see Fig.~\ref{fig4-pulsesrep}). These factors were not previously taken into account and,
as we will show, they do contribute in the entanglement dynamics.

The succession of pulses considered in this work is represented in Fig.~\ref{fig4-pulsesrep}. Each pulse
has a Gaussian envelope $f(t)$ and negative linear chirp, being described by an electric field $E(t)=E_0
f(t) \cos [\omega_L t + \varphi(t)]$, where $\omega_L/2\pi$ is the central frequency reached at $t=t_{P}$,
and $\varphi(t)$  a phase which is a quadratic function of time. The Gaussian envelope $f(t)=\sqrt{\tau_L/\tau_C}
\exp \lbrace -2 \ln 2 [(t-t_P)/\tau_C]^2 \rbrace$ is centered at $t=t_{P}$, having the temporal width $\tau_{C}$
defined as the full width at half maximum (FWHM) of the temporal intensity  profile  $E^2_0 f^2(t)$. The maximum
of $f(t)$ is at $f(t_P)=\sqrt{\tau_L/ \tau_C}$, where $\tau_{L}$ is the temporal width of the transform limited
pulse (before chirping). Such a pulse is characterized by several parameters belonging to the spectral and temporal
domains, which are carefully chosen in order to control the system evolution (a detailed analysis is contained in Refs.~\cite{vatasescu12,elianeepjd04}).

\begin{figure}
\includegraphics[width=0.9\columnwidth]{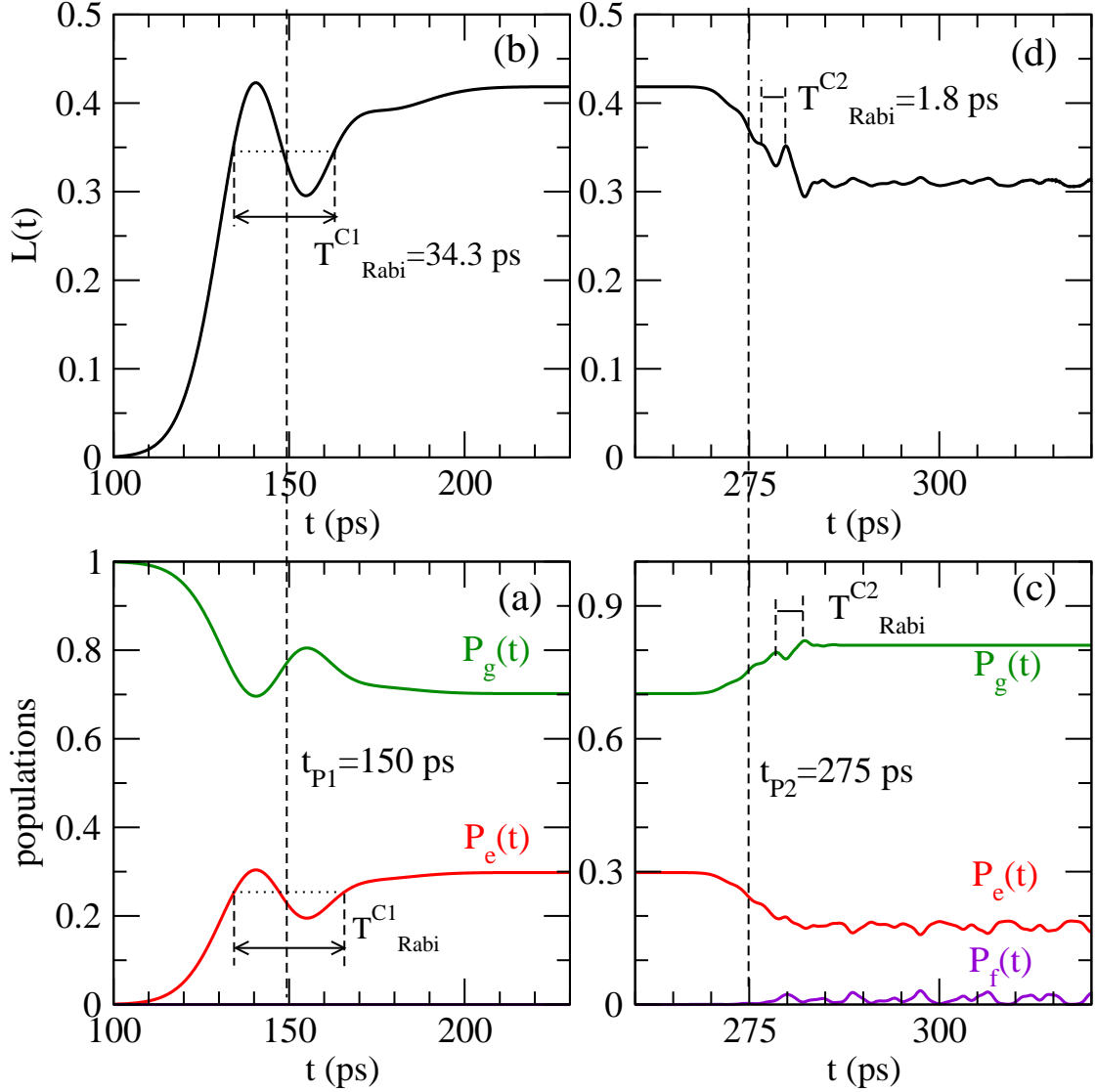}%
\caption{\label{fig5-normpurL1L2} (Color online) Time evolution of the linear entropy and electronic populations
during the first sequence of pulses (see Fig.~\ref{fig4-pulsesrep}). (a), (b) The population evolution 
in the three electronic channels, and the linear entropy evolution, respectively, during the first chirped pulse. (c), (d) The population evolution and linear entropy evolution, respectively, during the second chirped pulse.}
\end{figure}

The vibrational dynamics is obtained by solving numerically the time-dependent Schr\"odinger equation
associated with the radial motion of the vibrational wavepackets. For each pulse, the  coupling between
electronic surfaces is treated by using the rotating wave approximation with the  corresponding  carrier
frequency  $\omega_L/2\pi$. Then,  for a given pulse, the coupled equations for the evolution of the radial
wavefunctions $\Psi^{\omega}_{g,e,f}(R,t)$ in the dressed diabatic potentials $V^{\prime}_{e}(R)=V_{e}(R)$,
$V^{\prime}_{f}(R)=V_{f}(R)$, and $V^{\prime}_g(R)=V_g(R)+\hbar \omega_L$, can be written as
\begin{widetext}
\begin{eqnarray}
\label{tschreq}
i\hbar\frac{\partial}{\partial t}\left(\begin{array}{c}
 \Psi^{\omega}_{e}(R,t)\\
\Psi^{\omega}_{f}(R,t)\\
\Psi^{\omega}_{g}(R,t)
 \end{array}\right)=
\left(\begin{array}{ccc}
 {\bf \hat T} + V^{\prime}_{e}(R)  &
 V_{12}(R) &
-W_L f(t) e^{-i \varphi(t)} \\
V_{12} (R) &
 {\bf \hat T}+ V^{\prime}_{f}(R) &
0  \\
 -W_L f(t) e^{i \varphi(t)}  &
0  &
 {\bf \hat T} + V^{\prime}_g(R)
 \end{array} \right)
 \left( \begin{array}{c}
 \Psi^{\omega}_{e}(R,t)\\
\Psi^{\omega}_{f}(R,t) \\
\Psi^{\omega}_{g}(R,t)
 \end{array} \right).
\label{eqS3states}
\end{eqnarray}
\end{widetext}
Similar to the example analyzed in Sec.~\ref{sec:dyn1g}, ${\bf \hat T}$ is the kinetic energy operator
and  $W_L= -{\cal {E}}_0 D_{ge}/2$ is the laser coupling determined by the laser intensity $I$ (${\cal
{E}}_0=\sqrt{2I / c\epsilon_0}$) and the transition dipole moment $D_{ge}/2$. The Cs$_2$ molecular
potential curves used in the present work were described in Refs.~\cite{vatasescu06,vatasescu00}. The
non-adiabatic coupling between $0_g^-(6s,6p_{3/2})$ and $0_g^-(6s,5d)$ electronic potentials is modeled 
using a radial coupling $V_{12}(R)$ of Gaussian form \cite{vatasescu06} in the Hamiltonian matrix of
Eq.~(\ref{eqS3states}). The numerical methods used to solve Eq.~(\ref{eqS3states}) are those already
mentioned in Sec.~\ref{sec:dyn1g}.

\begin{figure}
\includegraphics[width=0.9\columnwidth]{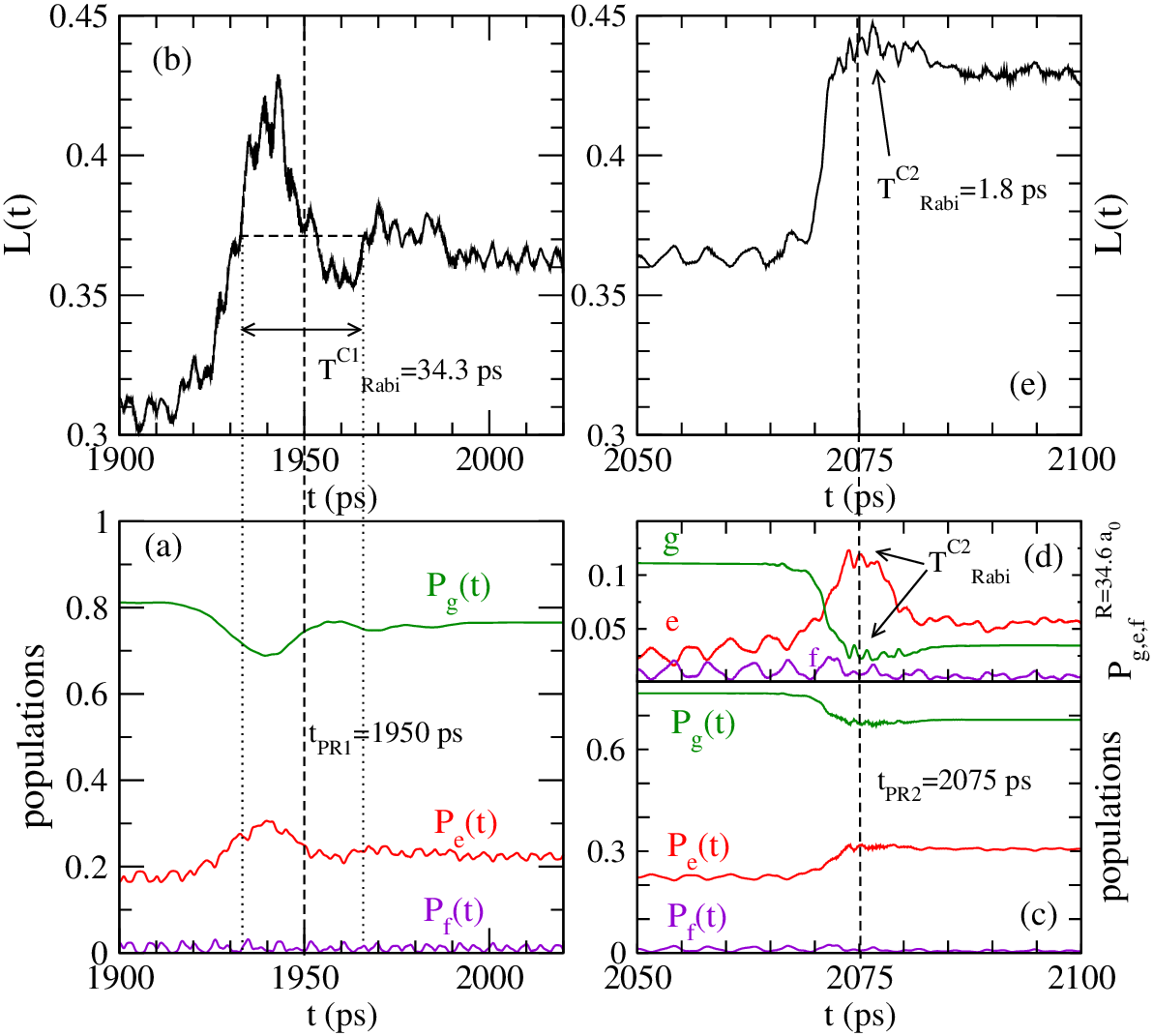}%
\caption{\label{fig6-normpurL1L2R} (Color online) Time evolution of the linear entropy and  electronic
populations during the repetition of the pulse sequence after 1800 ps (see Fig.~\ref{fig4-pulsesrep}).
(a),(b) The population evolution in the three electronic channels, and the linear entropy evolution,
respectively, during the repetition of the first chirped pulse. (c) Population evolution during the
repetition of the second chirped pulse. (d) Evolution of the partial populations $P_{g,e,f}^{R=34.6 a_0}
=\int^{R=34.6 a_0} | \Psi_{g,e,f}(R',t) |^{2} dR'$ during the repetition of the second chirped pulse. (e)
Evolution of the linear entropy during the repetition of the second chirped pulse.}
\end{figure}

The populations $P_{g}(t)$, $P_{e}(t)$, $P_{f}(t)$ in each electronic state are calculated
from the vibrational wavepackets as
\begin{equation}
 P_{g,e,f}(t) = \int^{L_R} | \Psi^{\omega}_{g,e,f}(R',t) |^{2} dR',
\end{equation}
where $L_R=1060$ a$_0$ is the length of the spatial grid used to solve Eq.~(\ref{eqS3states}) by
wavepackets propagation. The total population is normalized at 1 on the spatial grid ($P_g(t)+P_{e}
(t)+P_{f}(t)=1$), with $P_g(t=0)=1$.

\begin{figure}
\includegraphics[width=0.9\columnwidth]{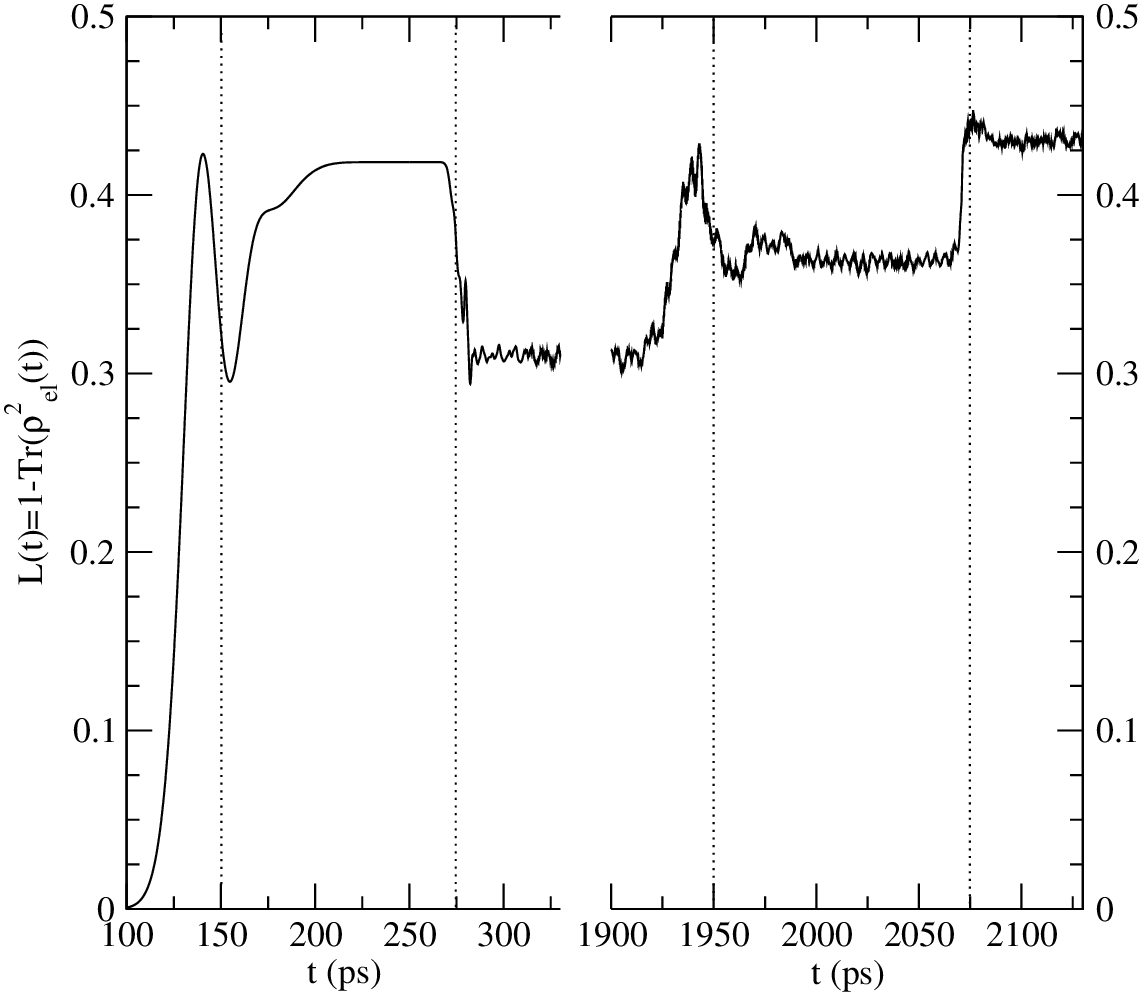}%
\caption{\label{fig7-linentrop3P} Dynamics of entanglement: time evolution of the linear entropy $L(t)=1-Tr
(\hat{\rho}^2_{el}(t))$ during the pulse sequence represented in Fig. \ref{fig4-pulsesrep}. The vertical
dotted lines indicate the instants $t_{P1}$, $t_{P2}$,  $t_{PR1}$, $t_{PR2}$ corresponding to the
maximum of every pulse envelope, as it is shown in Fig. \ref{fig4-pulsesrep}.}
\end{figure}

The linear entropy, $L(t)=1-\text{Tr}(\hat{\rho}^2_{el}(t))$, calculated with Eq.~(\ref{purityred3}),
is used to characterize the entanglement dynamics during the whole process.  Figs.~\ref{fig5-normpurL1L2},\ref{fig6-normpurL1L2R} show the evolution of the
linear entropy and electronic populations during each pulse. In Fig.~\ref{fig7-linentrop3P}
is represented the overall linear entropy evolution. In the following, we will analyze these results.

In the population evolution appears the chirped Rabi period \cite{chirpPRA04}, which is 
a characteristic time associated with the action of a chirped pulse:
\begin{equation}
T^C_{Rabi}(t_P)=\sqrt{\frac{\tau_C}{\tau_L}}\frac{\hbar \pi}{W_L}.
\end{equation}

\begin{figure}
\includegraphics[width=0.9\columnwidth]{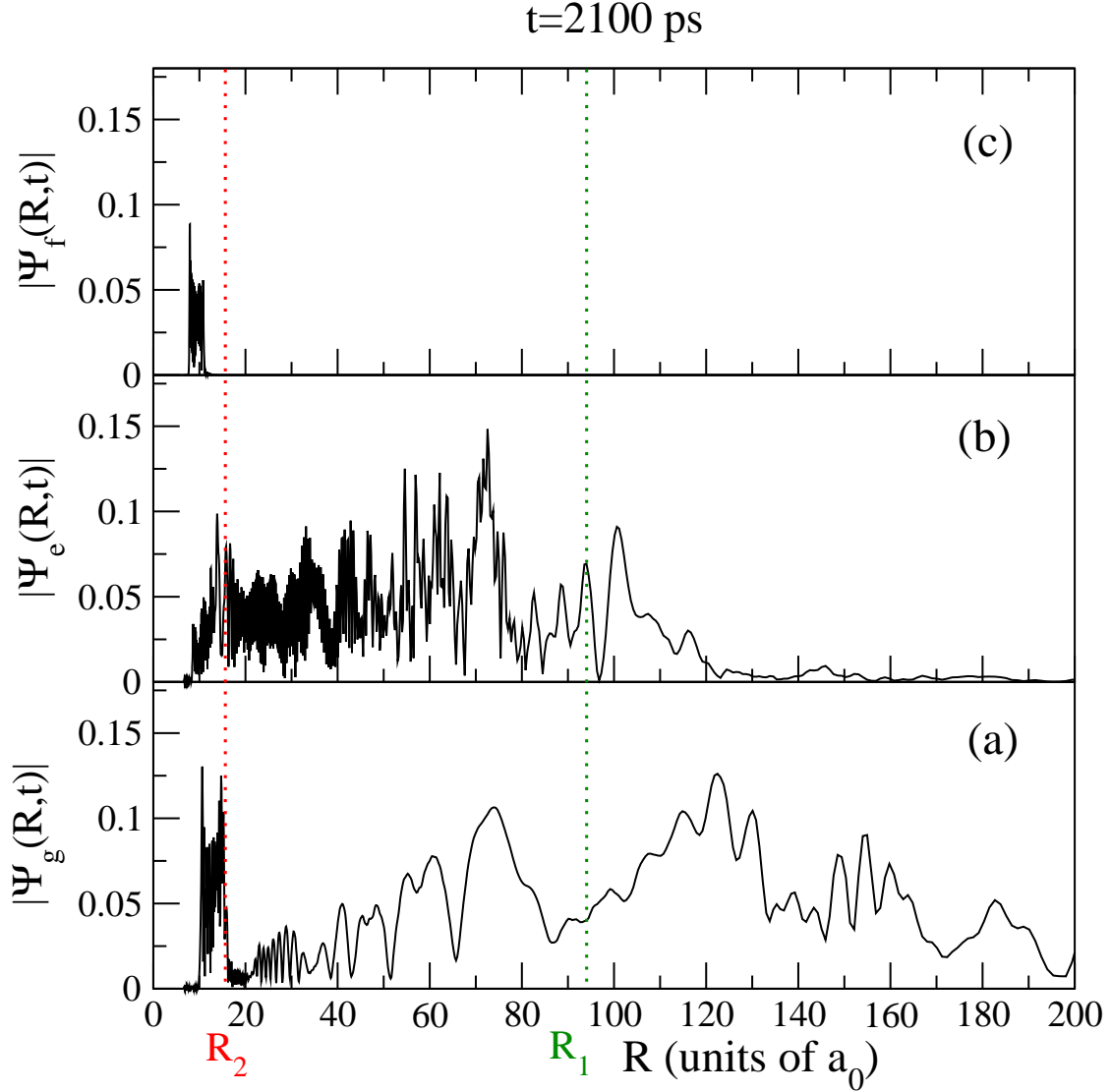}%
\caption{\label{fig8-wf3t2100} (Color online) Vibrational components $\psi_{g,e,f}(R,t)$ of the pure
entangled state $|\Psi_{el,vib}(t) > = |g> \bigotimes |\psi_g(t) > + |e> \bigotimes |\psi_e(t) > +
|f> \bigotimes |\psi_f(t) > $ created by the succesion of pulses at t=2100 ps. }
\end{figure}

The first chirped pulse, characterized by a coupling strength $W_{L1}=0.74$ cm$^{-1}$ and a chirped Rabi
period $T^{C1}_{Rabi}(t_{P1})=34.3$ ps, acts at large $R$ distances and produces a transfer of population from
the  $g$  electronic state to the $e$ state: the characteristic period $T^{C1}_{Rabi}(t_{P1})$ appears clearly in
the evolution of the populations and linear entropy, in Figs.~\ref{fig5-normpurL1L2}(a),(b). The
second pulse comes with a coupling strength $W_{L2}=24.69$ cm$^{-1}$ and a chirped Rabi period $T^{C2}_{Rabi}
(t_{P2})=1.8$ ps, and operates an exchange of populations at much smaller $R \le 35$ a$_0$ distances, bringing
population back to the ground electronic state $g$, but in strongly bound vibrational levels. Its characteristic
Rabi period $T^{C2}_{Rabi}(t_{P2})$ can be distinguished in the evolution of the populations and linear entropy,
in Figs.~\ref{fig5-normpurL1L2}(c,d), but with a smaller amplitude, due to the smaller amount of transferred
population. On the other hand, during the second pulse which brings population into the inner zone, the evolution
of the populations and linear entropy begins to show the beats due to the non-adiabatic coupling between the
$e$ and $f$ states.

The evolution during the repetition of the pulse sequence after 1800 ps is represented in Fig.~\ref{fig6-normpurL1L2R}(a,b) for the first pulse and in Fig.~\ref{fig6-normpurL1L2R}(c,d,e) for the second pulse. As the second pulse
operates at small distances $R$, it appears that to understand its effects one has to represent also the partial
populations $P_{g,e,f}^{R=34.6 a_0}=\int^{R=34.6 a_0} | \Psi_{g,e,f}(R',t) |^{2} dR'$, calculated by integrating
the vibrational wavepackets up to $R=34.6 a_0$ (Fig.~\ref{fig6-normpurL1L2R}(d)). Then, the chirped Rabi period
$T^{C2}_{Rabi}(t_{P2})$ may be easily identified in their evolution, as in the linear entropy evolution presented in Fig.~\ref{fig6-normpurL1L2R}(e).

The repetition of the first pulse feels the "void" left in the $g$ initial wavefunction by the initial first pulse
(see Fig.~\ref{fig8-wf3t2100}(a)), and as a result less population is transferred from the ground state $g$ to the
excited state $e$ at large distances (Fig.~\ref{fig6-normpurL1L2R}(a)). Nevertheless, by bringing closer the
electronic populations $P_g$ and $P_e$, the first pulse (the initial and its repetition) leads to an increase of
the linear entropy (Fig.~\ref{fig7-linentrop3P}).
On the other hand, the second pulse in the sequence
(which operates at small distances) has different effects initially and in its repetition.
In the first stage, it transfers population in low vibrational levels of the ground state $g$, which has a "purification effect" on the overall state,
lowering the linear entropy (see Figs.~\ref{fig5-normpurL1L2}(c,d) and Fig.~\ref{fig7-linentrop3P}). 
In contrast,
its repetition transfers population back in the inner well of the excited state $e$ (Fig.~\ref{fig6-normpurL1L2R}(c,d)),
bringing even closer the electronic populations and increasing $L(t)$ (Fig.~\ref{fig6-normpurL1L2R}(e)).
Therefore, it appears that the repetition of the pulse sequence is significant to the overall picture, which is
best seen in the evolution of the linear entropy during the process (Fig.~\ref{fig7-linentrop3P}).

The succession of pulses creates a final state $|\Psi_{el,vib}(t) >$ with significant entanglement, if we take
into account that the linear entropy is maximally bounded by $2/3$, and here $L(t)$ attains 0.42. The vibrational
components $\psi_{g,e,f}(R,t)$ of the pure entangled state $|\Psi_{el,vib}(t) >$, according to Eq.~(\ref{pure-elvib3}),
are shown in Fig.~\ref{fig8-wf3t2100} (other decompositions of $|\Psi_{el,vib}(t) >$ are equally possible, as in
Eq.~(\ref{pure-elvib3j})).

\section{\label{sec:conclu} Conclusion}

We have investigated the entanglement between electronic and nuclear degrees of freedom produced by
vibronic couplings in pure states of the Hilbert space $\cal{H}$$=$$\cal{H}$$_{el}$$\bigotimes$$
\cal{H}$$_{vib}$. Expressions for the von Neumann entanglement entropy and the reduced linear entropy were derived
for the  $2 \times N_v$ and $3 \times N_v$ cases of the bipartite entanglement (el $\bigotimes$ vib),
relating these entanglement measures to quantities specific to the intramolecular dynamics, such as
the electronic populations and the vibronic coherences.

The entanglement dynamics was analyzed in two
cases of laser coupling between electronic states, 
using as an example the Cs$_2$ molecule.
In the first case, treated in Sec.~\ref{sec:dyn1g}, we have simulated the vibrational dynamics for
two electronic states of the Cs$_2$ molecule, $a^3\Sigma_{u}^{+}(6s,6s)$ and $1_g(6s,6p_{3/2})$,
which are coupled by a laser pulse. We show that the Rabi period due to the vibronic
laser coupling is also a characteristic time in the evolution of the von Neumann entropy and of
the reduced linear entropy. The pulse creates the conditions for the equalization of population
between the two electronic channels, producing an electronically maximally entangled state in
several stages of the temporal evolution.

The second case, described in Sec.~\ref{sec:3elchirp}, is related to a theoretical control scheme
proposed to create Cs$_2$ vibrationally cold molecules using a multichannel tunneling in the
$0_g^-(6s,6p_{3/2})$ and $0_g^-(6s,5d)$ electronic states coupled through a non-adiabatic coupling
generated by the spin-orbit interaction. The scheme employs three electronic states ($a^3\Sigma_{u}^{+}
(6s,6s)$, $0_g^-(6s,6p_{3/2})$, and $0_g^-(6s,5d)$) coupled by a sequence of two chirped laser pulses.
In addition to previous treatments, we have introduced in the simulation of the dynamics the
non-adiabatic coupling between $0_g^-(6s,6p_{3/2})$ and $0_g^-(6s,5d)$ at short distances, and the
repetition of the pulse sequence. In these conditions we have analyzed the entanglement dynamics
quantified by the reduced linear entropy. The chirped Rabi period characteristic to each pulse can
be identified in the linear entropy evolution, as well as the beats period due to the non-adiabatic
radial coupling between the tunneling channels $0_g^-(6s,6p_{3/2})$ and $0_g^-(6s,5d)$.
We have shown that the repetition of the pulse sequence has considerable influence on the process,
diminishing the purification effect of the first sequence and increasing the entanglement in the
final state.

In both cases, the results show that the characteristic times related to the vibronic couplings 
and the vibrational motion are present in the entanglement structure.
The linear entropy, calculated from the purity of the electronic reduced density matrix, appears
as an interesting sensor for the correlations between the electronic channels, emphasizing
explicitly the role played by the vibronic coherences. This property  could qualify the linear
entropy as a useful reference in control schemes of the molecular coherence and entanglement
\cite{buchleitner13}.

In a molecule controlled by laser pulses which leave more than one electronic state populated,
electronic-vibrational entanglement is always produced. The amount of entanglement will depend
on the "entangling power" of the quantum evolution \cite{zanardi00}, which is directed  here by
the laser pulses, and could, in principle, be controlled.

Molecules are systems whose entanglement properties are beginning to be explored. Many phenomena
are expected to contribute to the intramolecular dynamics and they could be interrogated in future
developments regarding  electronic-nuclear entanglement in isolated molecules: electronic energy
relaxation, vibrational energy redistribution and relaxation, and various coupling mechanisms
\cite{bookJungen}. We hope that the present work will help in these possible developments, and also
in future studies aiming to investigate the environment effects and the control of entanglement
in molecules.

\begin{acknowledgments}
I am grateful to Ovidiu Patu for the critical reading of the manuscript.
This work was supported by the LAPLAS 3 39N Research Program of the Romanian Ministry of Education
and Research.
\end{acknowledgments}

\bibliography{artMV-entang}

\end{document}